\documentclass[aps,prb,reprint,showpacs,floatfix,citeautoscript,longbibliography,superscriptaddress]{revtex4-1}
\usepackage{graphicx}
\usepackage{bm,amsmath,amssymb,mathrsfs,dcolumn}
\usepackage[colorlinks=true, linkcolor=blue, citecolor=blue, urlcolor=blue, linktoc=page, bookmarks=false, pdfstartview={FitH}, pdfborder={0 0 0.0 [3 3]}]{hyperref} 
\usepackage[usenames]{color} 
\hypersetup{pdfborder=0 0 0,colorlinks=true,citecolor=blue,linkcolor=blue}
\usepackage[final]{pdfpages}

\begin{document}
\title{Spin-orbit-coupling induced torque in ballistic domain walls: equivalence of charge-pumping and nonequilibrium magnetization formalisms}
\author{Zhe Yuan}
\affiliation{The Center for Advanced Quantum Studies and Department of Physics, Beijing Normal University, 100875 Beijing, China}
\affiliation{Faculty of Science and Technology and MESA$^+$ Institute for Nanotechnology, University of Twente, P.O. Box 217, 7500 AE Enschede, The Netherlands}
\affiliation{Institut f{\"u}r Physik, Johannes Gutenberg--Universit{\"a}t Mainz, 55128 Mainz, Germany}
\author{Paul J. Kelly}
\affiliation{Faculty of Science and Technology and MESA$^+$ Institute for Nanotechnology, University of Twente, P.O. Box 217, 7500 AE Enschede, The Netherlands}
\date{\today}
\begin{abstract}
To study the effect of spin-orbit coupling (SOC) on spin-transfer torque in magnetic materials, we have implemented two theoretical formalisms that can accommodate SOC. Using the ``charge-pumping'' formalism, we find two contributions to the out-of-plane spin-transfer torque parameter $\beta$ in ballistic Ni domain walls (DWs). For short DWs, the nonadiabatic reflection of conduction electrons caused by the rapid spatial variation of the exchange potential results in an out-of-plane torque that increases rapidly with decreasing DW length. For long DWs, the Fermi level conduction channel anisotropy that gives rise to an intrinsic DW resistance in the presence of SOC leads to a linear dependence of $\beta$ on the DW length. To understand this counterintuitive divergence of $\beta$ in the long DW limit, we use the ``nonequilibrium magnetization'' formalism to examine the spatially resolved spin-transfer torque. The SOC-induced out-of-plane torque in ballistic DWs is found to be quantitatively consistent with the values obtained using the charge-pumping calculations indicating the equivalence of the two theoretical methods.
\end{abstract}
\maketitle

\section{Introduction}
An electron has an intrinsic (spin) angular momentum and associated with this a magnetic moment. When an electric current flows, it is accompanied by a flow of spin angular momentum. For non-magnetic (NM) materials like copper, the current of electrons with spin in a particular direction (e.g. ``up'') is compensated by an equal current of electrons with spin in the opposite direction (``down'') so there is no net flow of spin angular momentum. In a ferromagnetic (FM) material with an unequal number of spin-up and spin-down electrons, there is a flow of spin angular momentum but this only has observable consequences when translational symmetry is broken. This happens, for example, at an interface with a NM metal where spin angular moment is injected into the NM metal leading to ``spin accumulation'' \cite{Yang:natp08}. It also happens when the magnetization direction depends on the position in space as in a domain wall (DW) where there is a continuous transition of the magnetization direction between two domains in which it is entirely collinear (up or down for the $180^\circ$ DW sketched in Fig.~\ref{fig:1}). In this case, spin angular momentum is transported by an electric current from one region of space to another where it leads to an imbalance and tends to realign the angular momentum and magnetization direction of both regions of space. This phenomenon is called ``spin-transfer torque'' (STT) \cite{Slonczewski:jmmm96, Berger:prb96, Brataas:natm12} and it forms the basis for writing information in magnetic random access memories \cite{Akerman:sc05, Kryder:ieeem09} or for microwave frequency STT oscillators where the injected spin forces a magnetization to precess with GHz frequency \cite{Berger:prb96, Ruotolo:natn09, Slavin:natn09}. Passage of a spin-polarized current can also cause a domain wall to move. This is the principle behind a new form of shift register called ``racetrack memory'' \cite{Parkin:sc08, Franken:natn12}.

The STT was first predicted based upon the conservation of spin angular momentum; a loss of spin current, ${\bf \nabla \cdot  j}_s$, corresponds to a torque $- d{\bf s}/dt$ exerted on the local magnetization\cite{Slonczewski:jmmm96, Berger:prb96}. Various theoretical methods were proposed to compute STTs with realistic electronic structures \cite{Waintal:prb00, Bauer:mseb01, Haney:prb07, Wang:prb08} and a number of these were implemented with first-principles electronic structure calculations \cite{Xia:prb02, Zwierzycki:prb05, Haney:prb07, Wang:prb08}. Not all are suitable for studying the effect of spin-orbit coupling (SOC) on STT though \cite{Haney:prl10}. The spin-orbit interaction couples the electron spin to its orbital motion and the STT exerted on a local magnetization can be larger than the maximum spin angular momentum that can be transferred from conduction electrons, i.e. an amount of $\hbar$ per electron. It was recently found that STTs arising from SOC can be more efficient in driving magnetization switching, forcing oscillation, or moving magnetic DWs \cite{Brataas:natn14}.

Two quite distinct theoretical formalisms have been proposed to calculate the STT without assuming spin angular momentum conservation. The method proposed by the Austin group \cite{Nunez:ssc06} is to calculate the STT in terms of the exchange interaction between the local magnetization and a nonequilibrium magnetization generated by the current. We will refer to this as the nonequilibrium magnetization (NEM) scheme. The effect of SOC is explicitly included in the Hamiltonian that is used to determine the current-induced nonequilibrium magnetization. The NEM scheme has been applied to calculate STT in spin valves \cite{Haney:prb07}, magnetic tunnel junctions \cite{Heiliger:prl08}, and ferromagnet$|$normal metal bilayers. \cite{Haney:prb13, Freimuth:prb14, Freimuth:prb15} The other method is to consider the charge current pumped by a time varying magnetization. By making use of Onsager reciprocity relations, this can be used to derive the STT \cite{Hals:prl09, Hals:epl10}. The charge pumping formalism is also applicable when SOC is included in the Hamiltonian. 

In this paper, we study the out-of-plane STT in ballistic DWs taking nickel as an example. While numerical values of these torques have been reported in the diffusive regime for real materials using realistic electronic structures \cite{Gilmore:prb11, Brataas:12}, its physical origin remains unclear. \cite{Akosa:prb15} Experimental observations are usually interpreted by comparing the measured velocities of current-driven DWs with the results of micromagnetic simulations, a procedure that is not straightforward. For instance, the Gilbert damping must be accurately taken into account in simulations \cite{Weindler:prl14} but its form in noncollinear magnetizations is still the subject of discussion \cite{Yuan:prl14}. By implementing both the charge pumping and the NEM formalisms using first-principles scattering theory \cite{Xia:prb06, Starikov:prl10} and using them to calculate the out-of-plane STT in ballistic DWs, we demonstrate the quantitative equivalence of the two computational schemes. The STTs obtained in ballistic DWs can be understood in terms of the scattering of electrons by the noncollinear magnetization as characterized by the DW resistance (DWR). For very short DWs, the nonadiabatic reflection of conduction electrons by the large magnetization gradients gives rise to a relatively large DWR and out-of-plane torque. In the long DW limit, the DWR is dominated by the length-independent intrinsic DWR \cite{Nguyen:prl06,  Oszwaldowski:prb06} that results from an anisotropy in the distribution of conducting channels induced by SOC in combination with the noncollinear magnetization. Electron reflection due to the intrinsic DWR gives rise to an out-of-plane torque that scales linearly with the DW length. We calculate the amplitude of this torque using the two different methods and obtain values in good quantitative agreement with each other.

The rest of this paper is organized as follows. The charge-pumping formalism is outlined in Sec.~\ref{sec:pumping} followed by the results of calculations for the DWR and out-of-plane STT parameter. We present the details of the NEM scheme in Sec.~\ref{sec:nems} and verify the implementation for a spin valve system benchmarked in the literature; in the absence of SOC we can compare to STTs calculated using spin conservation \cite{Haney:prb07, Wang:prb08}. In Sec.~\ref{sec:nickel}, we use the NEM scheme to calculate spatially resolved STTs in ballistic nickel DWs. The out-of-plane component quantitatively agrees with the values obtained in Sec.~\ref{sec:pumping} using the charge-pumping formalism. Some conclusions are drawn in Sec.~\ref{sec:conclusion}.

\section{Charge-pumping formalism\label{sec:pumping}}
\subsection{Formalism}
In this section, we briefly outline the charge-pumping formalism of Ref.~\onlinecite{Hals:prl09} and how it can be combined with first-principles scattering calculations that include SOC. In the presence of an electrical current ${\bf j}$ with spin polarization $P$, the dynamics of a magnetization ${\bf M}({\bf r})$ with magnitude $M_s$ and direction $\hat{\mathbf M}$ is described by the phenomenological generalized Landau-Lifshitz-Gilbert (LLG) equation \cite{Zhang:prl04, Thiaville:epl05, Tserkovnyak:prb06, Tserkovnyak:jmmm08, Brataas:natm12}
\begin{equation}
\frac{d \hat{\mathbf M}}{d t}=-\gamma\hat{\mathbf M}\times\mathbf H_{\mathrm{eff}}+\alpha\hat{\mathbf M}\times\frac{d \hat{\mathbf M}}{d t}-\left(1-\beta\hat{\mathbf M} \times\right)\left(\mathbf v_s\cdot\nabla\right)\hat{\mathbf M}, \label{eq:llg}
\end{equation}
where $\mathbf H_{\mathrm{eff}}$ is the effective magnetic field, $\gamma=g\mu_B/\hbar$ is the gyromagnetic ratio expressed in terms of the Land{\'e} $g$ factor and Bohr magneton $\mu_B$, and $\mathbf v_s=g\mu_B P\mathbf j/(2eM_s)$ is an effective velocity. In this paper, we use the following conventions. Electrons flow from the left to the right lead along $\hat z$ and the charge current density $\mathbf j=-\vert j\vert\hat z$. The electron charge is negative, $e=-\vert e\vert$. The current polarization $P$ in ballistic Ni is found to be negative since the minority-spin electrons have a larger state density at the Fermi energy than the majority-spin electrons and contribute more to the Sharvin conductance. 

In Eq.~(\ref{eq:llg}), $\alpha$ and $\beta$ are material-dependent constants used to characterize the Gilbert damping and the out-of-plane STT respectively. The Gilbert damping in a DW depends not only on the magnetization gradient but also on the particular mode of precession and can be calculated using first-principles scattering theory \cite{Brataas:prb11, Yuan:prl14}. The out-of-plane STT parameter $\beta$ plays an important role in current-driven DW motion and is the key quantity that we calculate in this paper. We consider Walker-profile \cite{Schryer:jap74} Bloch DWs with magnetization profile $\hat{\mathbf M}(z)=(-\tanh\frac{z-r_w}{\lambda_w},-\mathrm{sech}\frac{z-r_w}{\lambda_w},0)$, or N{\'e}el DWs with profile $\hat{\mathbf M}(z)=(\mathrm{sech}\frac{z-r_w}{\lambda_w},0,\tanh\frac{z-r_w}{\lambda_w})$ that are centered at $r_w$ and have lengths $\lambda_w$ \cite{Schryer:jap74}. If the DW is displaced rigidly so that the magnetization varies in time only via the DW center, i.e. 
\begin{equation}
\frac{d\hat{\mathbf M}}{dt}=\dot{r}_w \frac{d\hat{\mathbf M}}{\,\, dr_w},
\end{equation}
the DW profile can be explicitly substituted into Eq.~(\ref{eq:llg}) to find a solution of the generalized LLG equation \cite{Tserkovnyak:jmmm08},
\begin{equation}
\dot r_w=\frac{\gamma\lambda_w}{\alpha}H_{\rm ext}+\frac{\hbar\gamma P\beta G}{2eAM_s\alpha}V.\label{eq:rw}
\end{equation}
Equation~(\ref{eq:rw}) describes the low current density steady state motion of a DW in response to an external field $H_{\rm ext}$ and electrical voltage $V$. Here $G$ is the conductance of the DW and $A$ is the cross sectional area. 

The process reciprocal to current- or bias-driven DW motion is the ``pumping'' of a charge current by a moving DW \cite{Hals:prl09, Hals:epl10, Duine:prb09}. These  (reciprocal) processes can be described using coupled thermodynamic equations. We first identify two thermodynamical fluxes, the DW velocity $\dot{r}_w$ and charge current $I$. The conjugate forces defined by the requirement that the energy dissipation is given by the product of the flux and its conjugate force \cite{deGroot:52} are found to be $2AM_sH_{\rm ext}$ and $V$, respectively. The coupled equations can then be written as
\begin{eqnarray}
\left(\begin{array}{c}\dot{r}_w \\ I\end{array}\right)=\left(\begin{array}{cc}\mathcal L_{11} & \mathcal L_{12} \\ \mathcal L_{21} & \mathcal L_{22} \end{array}\right)\left(\begin{array}{c}2AM_sH_{\rm ext} \\ V\end{array}\right),\label{eq:coupled}
\end{eqnarray}
and the coefficient $\mathcal L_{ij}$ characterizing how the $i$-th flux is induced by the $j$-th force can be derived as follows. Comparison of the first line of Eq.~(\ref{eq:coupled}) with Eq.~(\ref{eq:rw}) yields the coefficients $\mathcal L_{11}=\gamma\lambda_w/(2AM_s \alpha)$  and $\mathcal L_{12}=\hbar\gamma P\beta G/(2eAM_s\alpha)$. According to Ohm's law $\mathcal L_{22}$ is just the conductance $G$ of the DW. Reciprocity of the Onsager relations makes it possible to determine the last unknown coefficient $\mathcal L_{21}=\mathcal L_{12}$. Knowing all the coefficients in Eq.~(\ref{eq:coupled}), the charge pumped by a DW forced to move by an external magnetic field $H_{\rm ext}$ is found to be
\begin{eqnarray}
I=\mathcal L_{21}2AM_sH_{\rm ext}=\frac{\mathcal L_{21}}{\mathcal L_{11}}\dot{r}_w =\frac{\hbar P\beta G}{e\lambda_w}\dot r_w.
\label{eq:pump1}
\end{eqnarray}
Using the concept of parametric pumping \cite{Tserkovnyak:rmp05}, the electrical  current induced by a moving DW can alternatively be expressed in terms of the scattering matrix of the system, $\mathbf S=\left(\begin{array}{cc} \mathbf r & \mathbf t' \\ \mathbf t & \mathbf r'\end{array}\right)$, as 
\begin{equation}
I=\frac{e\dot r_w}{4\pi}\mathrm{Im}\left[\mathrm{Tr}\left(\frac{\partial \mathbf S}{\partial r_w} \mathbf S^{\dagger} \bm{\Sigma} \right) \right],\label{eq:pump2}
\end{equation}
 with $\mathbf r(\mathbf r')$ and $\mathbf t(\mathbf t')$ comprising matrices of reflection and transmission amplitudes for states incident from left (right) leads, respectively. The matrix $\bm{\Sigma}=\left(\begin{array}{cc} {\bf 1} & {\bf 0} \\ {\bf 0} & -{\bf 1'} \end{array}\right)$ consists of the unit matrices $\bf 1$ and $\bf 1'$ that have the same dimensions as $\mathbf r$ and $\mathbf r'$, respectively. 

Comparing equations (\ref{eq:pump1}) and (\ref{eq:pump2}), and writing the conductance in terms of the transmission matrix $\mathbf t$ as
\begin{eqnarray}
G=\frac{e^2}{h}\mathrm{Tr}\left(\mathbf t\mathbf t^\dagger\right),
\label{eq:cond}
\end{eqnarray} 
we arrive at the required expression for $\beta$
\begin{equation}
\beta=\frac{\lambda_w}{2P\mathrm{Tr}\left(\mathbf t\mathbf t^{\dagger}\right)}\mathrm{Im}\left[\mathrm{Tr}\left(\frac{\partial \mathbf S}{\partial r_w} \mathbf S^{\dagger} \bm{\Sigma} \right) \right].
\label{eq:beta}
\end{equation}
Equations~(\ref{eq:cond}) and (\ref{eq:beta}) are used in this work to directly calculate the conductance (resistance) and the out-of-plane STT parameter $\beta$, respectively.

\subsection{Numerical details\label{sec:numerical}}
Our starting point is the electronic structure of bulk face-centered cubic (fcc) nickel calculated with tight-binding linearized muffin-tin orbitals (TB-LMTOs) \cite{Andersen:prb75, Andersen:prb86} within the framework of density functional theory. We use the local density approximation, specifically the exchange-correlation functional parameterized by von Barth and Hedin \cite{vonBarth:jpc72}, a minimal basis consisting of 9 orbitals ($s$, $p$ and $d$) per spin, and sample the first Brillouin zone of the fcc lattice with $120^3$ $k$ points. With the experimental lattice constant of 0.352~nm, the charge and spin densities of collinearly magnetized fcc nickel are calculated self-consistently within the atomic spheres approximation (ASA) to obtain a magnetic moment of 0.639 $\mu_B$ per nickel atom \cite{Andersen:85}. SOC is omitted in the self-consistent calculation since it is much smaller in energy than the band width and exchange interaction. 

\begin{figure}[t]
\includegraphics[width=1\columnwidth]{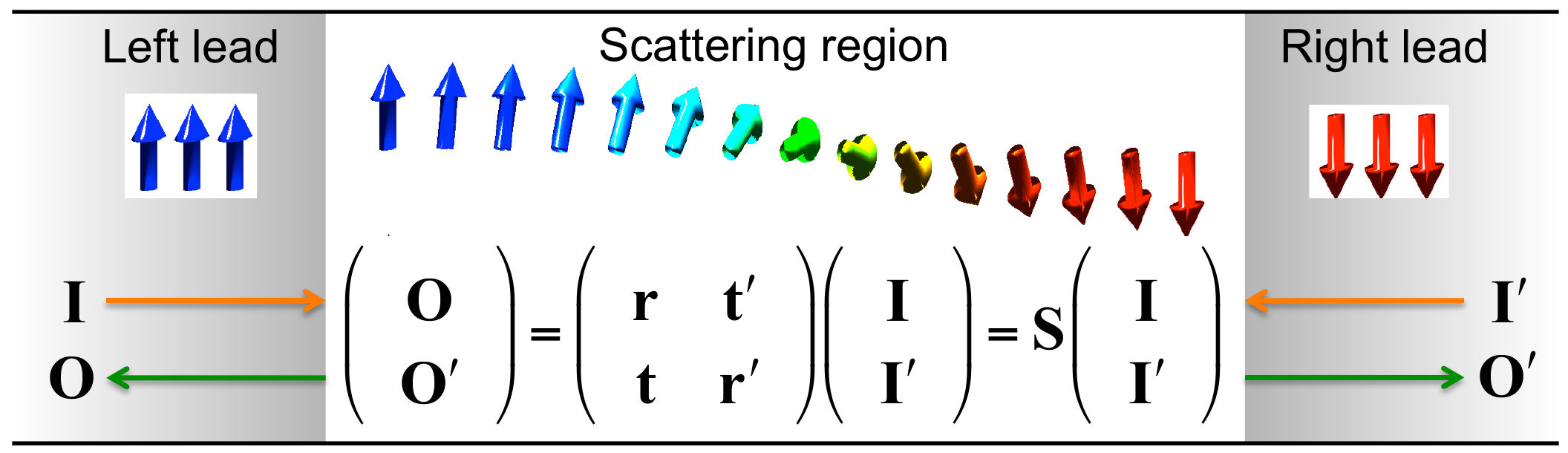}
\caption{Schematic illustration of the scattering theory. The scattering region consists of a 180$^\circ$ Walker-profile DW sandwiched between semiinfinite collinearly magnetized crystalline leads. Incoming, $\mathbf I$ ($\mathbf I'$), and outgoing, $\mathbf O$ ($\mathbf O'$), states in the left (right) lead are connected by the scattering matrix $\mathbf S$ which is made up of reflection $\mathbf r$ ($\mathbf r'$) and transmission $\mathbf t$ ($\mathbf t'$) matrices.}\label{fig:1}
\end{figure}
This electronic structure is appropriate for the semiinfinite leads. The scattering region also consists of perfectly crystalline nickel and purely for convenience we choose the transport direction to be along the fcc [111]. The electronic structure of the scattering region is constructed by rotating the bulk atomic sphere potentials in spin space so that the local quantization axis for every atomic sphere follows the Walker magnetization profile; see Fig.~\ref{fig:1}. 

We then consider the fate of each flux-normalized state $\psi_\mu^{I}(\mathbf k_\|;E_F)$ at the Fermi energy incident from the left lead. The transmitted and reflected wave functions far away from the scattering region can be expanded in terms of all possible outgoing propagating states in the right and left leads as $\sum_{\nu,\mathbf k'_\|}t_{\nu\mu}(\mathbf k'_\|,\mathbf k_\|)\psi_\nu^{O'}(\mathbf k'_\|;E_F)$ and $\sum_{\nu,\mathbf k'_\|}r_{\nu\mu}(\mathbf k'_\|,\mathbf k_\|)\psi_\nu^{O}(\mathbf k'_\|;E_F)$, respectively. The reflection and transmission coefficients $r_{\nu\mu}(\mathbf k'_\|,\mathbf k_\|)$ and $t_{\nu\mu}(\mathbf k'_\|,\mathbf k_\|)$ are determined using a ``wave-function matching'' scheme \cite{Ando:prb91} also implemented with TB-LMTOs \cite{Xia:prb06}. The same can be done for all states incident from the right lead to calculate $r'_{\nu\mu}(\mathbf k'_\|,\mathbf k_\|)$ and $t'_{\nu\mu}(\mathbf k'_\|,\mathbf k_\|)$ and so obtain the full scattering matrix $\mathbf S$ explicitly. 

In the absence of any disorder breaking the translational symmetry perpendicular to the transport direction, the parallel component $\mathbf k_\|$ of the bulk Bloch wavevector $\bf k$ is conserved and $S_{\nu\mu}(\mathbf k'_\|,\mathbf k_\|)=S_{\nu\mu}(\mathbf k_\|)\delta_{\mathbf k'_\|,\mathbf k_\|}$ for ``ballistic'' DWs. (Otherwise we could use a ``lateral supercell'' scheme to model disorder and allow transitions from one $\mathbf k_\|$ to another \cite{Xia:prb06, Liu:prb15}. It turns out that the calculated transport properties usually converge very quickly with respect to the size of the lateral supercell.) SOC is included in the transport calculations by using a Pauli Hamiltonian \cite{Daalderop:prb90a, Starikov:prl10}. Unless otherwise stated, the two-dimensional Brillouin zone is sampled using 600$\times$600 $k$ points to guarantee the convergence of the calculated conductance and out-of-plane STT parameter $\beta$.

\subsection{Domain-wall resistance}

\begin{figure}[b]
\includegraphics[width=0.95\columnwidth]{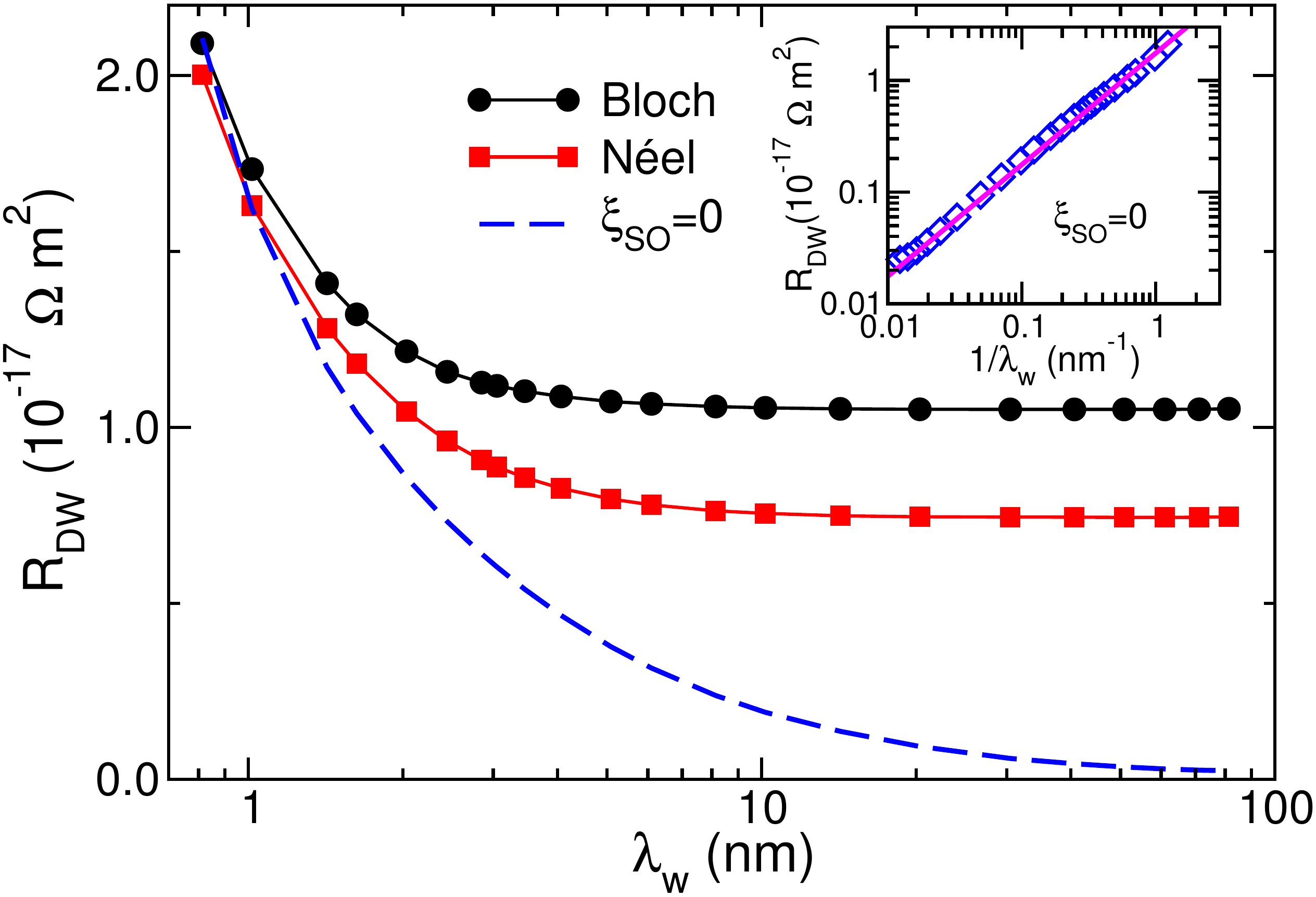}
\caption{DW resistance $R_{\rm DW}$ calculated for clean fcc Ni as a function of the DW length $\lambda_w$. For Bloch (black circles) and N{\'e}el (red squares) DWs, there are contributions to the DWR from (i) the nonadiabatic reflection of conduction electrons from short DWs ($\lambda_w<10$~nm) that decreases monotonically and vanishes in the long DW limit and (ii) the conduction channel mismatch in the presence of SOC that leads to a finite saturated DWR in the long DW limit. Without SOC, there is no distinction between Bloch and N{\'e}el DWs and only the nonadiabatic contribution is seen (dashed blue line). Inset: DWR without SOC replotted as a function of $1/\lambda_w$. The solid line illustrates the linear dependence.}\label{fig:2}
\end{figure}

Before calculating $\beta$ using Eq.~(\ref{eq:beta}), it is instructive to understand how electrons are scattered by a ballistic DW and to characterize this by the DW resistance (DWR) $R_{\rm DW}=1/G-1/G_0$, where $G$ and $G_0$ are the conductances of a DW and of a bulk metal with the saturation magnetization, respectively. In particular, $G_0$ is the Sharvin conductance of a bulk ballistic system \cite{Datta:95}. The DWR calculated for nickel is plotted in Fig.~\ref{fig:2} as a function of the DW length $\lambda_w$. Without SOC, $R_{\rm DW}$ is large for small values of $\lambda_w$ because the gradient of the local magnetization is large and the conduction electrons cannot follow the rapid variation of the effective potential \cite{Cabrera:pssb74a}. This nonadiabatic contribution to the DWR decreases monotonically with increasing DW length (dashed blue line) and vanishes in the long (adiabatic) limit in agreement with results found in earlier calculations\cite{vanHoof:prb99, Kudrnovsky:ss01}. In particular, the DWR for ballistic Ni without SOC is inversely proportional to the DW length as replotted in the inset to Fig.~\ref{fig:2}.

\begin{figure}[t]
\includegraphics[width=0.95\columnwidth]{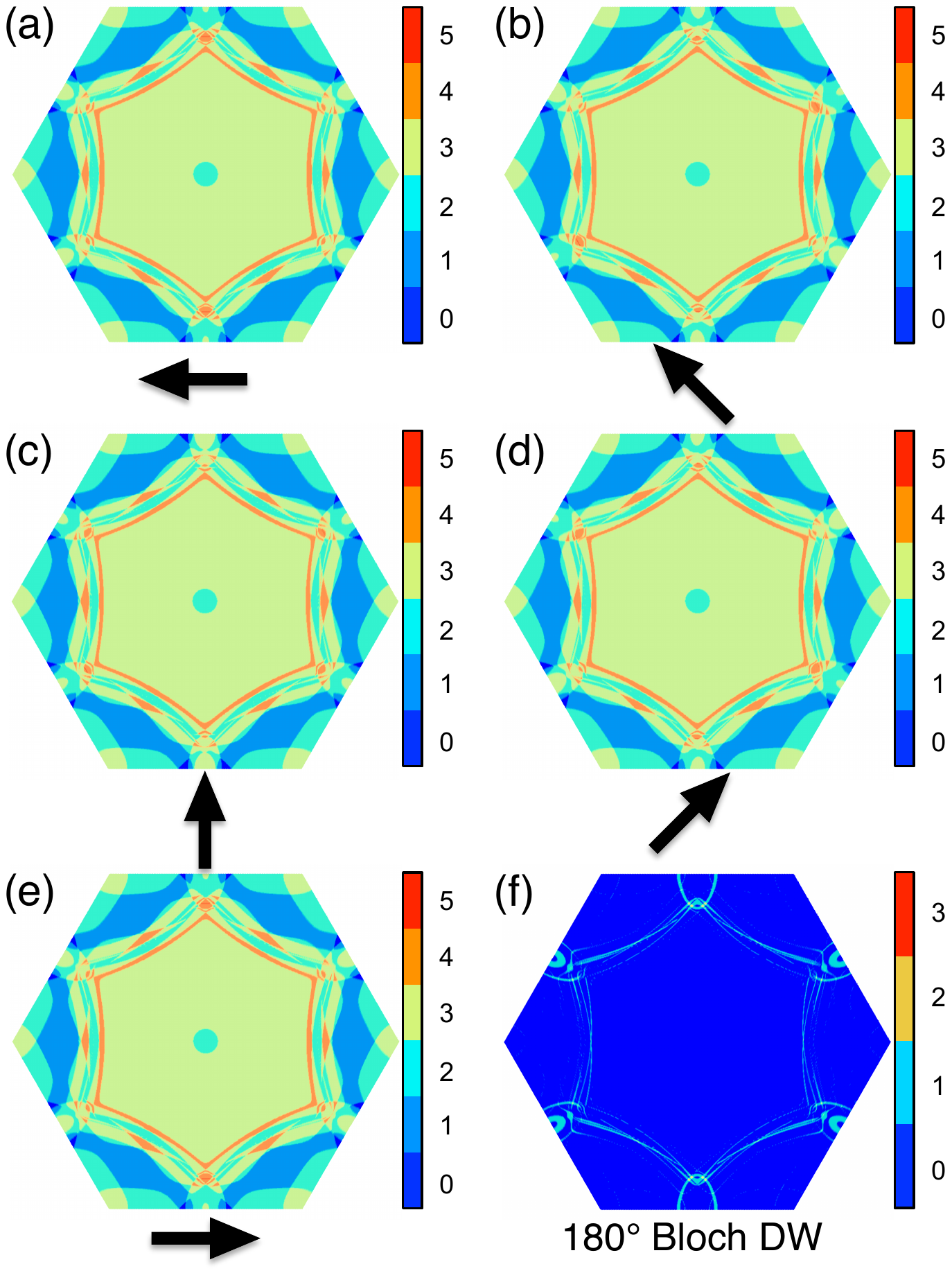}
\caption{(a)--(e) Calculated conduction channels at the Fermi level for fcc Ni along the [111] for the different magnetization orientations indicated by the arrows at the bottom of each panel. (f) The reflection probability of conduction electrons in a very long 180$^\circ$ Bloch DW. Large reflection probabilities are found for values of ${\bf k}_\|$ where the number of conduction channels depends upon the magnetization direction in (a)--(e). }\label{fig:3}
\end{figure} 
 
With SOC included, the DWR for small values of $\lambda_w$ is still dominated by the nonadiabatic contribution for both Bloch and N{\'e}el DWs while saturating to a finite value in the adiabatic, large $\lambda_w$ limit corresponding to the so-called intrinsic DWR \cite{Nguyen:prl06, Oszwaldowski:prb06}. It results from a variation in the number of conduction channels at the Fermi level on rotating the magnetization direction. Figs.~\ref{fig:3}(a)--(e) show the number of conduction channels in the first Brillouin zone in the [111] direction for different values of the magnetization direction of bulk Ni as a function of $\bf k_\|$, the component of the crystal momentum perpendicular to [111]. It is equivalent to the projection of the Fermi surface onto the two-dimensional plane perpendicular to the transport direction \cite{Schep:prl95, *Schep:prb98, Xu:prl06}. In ballistic systems, the crystal momentum of a propagating state is conserved and only the propagating channels that survive for all magnetization directions contribute to the total conductance. At some $\mathbf k_\|$ points the number of channels decreases as the magnetization rotates resulting in the reflection of the corresponding propagating electronic states. The total reflection in a long Bloch DW is plotted in Fig.~\ref{fig:3}(f). Large values of reflection probability are found for $\mathbf k_\|$ points where the number of conduction channels varies strongly with the magnetization direction shown in Figs.~\ref{fig:3}(a)--(e). Indeed, the intrinsic, saturated DWRs for Bloch and N{\'e}el walls can be well reproduced by counting the number of common conducting channels through the DWs. Since SOC is very weak in 3$d$ transition metals, it only slightly modifies their Fermi surfaces and the number of conduction channels for different magnetization orientations. Quantitatively, the intrinsic DWR is only 1.8\% and 1.3\% of the corresponding Sharvin resistance for the Bloch and N{\'e}el DWs, respectively. Note that the intrinsic DWR that is a nonlocal effect is eliminated in the diffusive regime, where spin-flip scattering and anisotropic magnetoresistance become the main mechanisms responsible for the DWR found there \cite{Yuan:prl12}.

\subsection{Out-of-plane spin-transfer torque parameter $\beta$}

\begin{figure}[b]
\includegraphics[width=0.95\columnwidth]{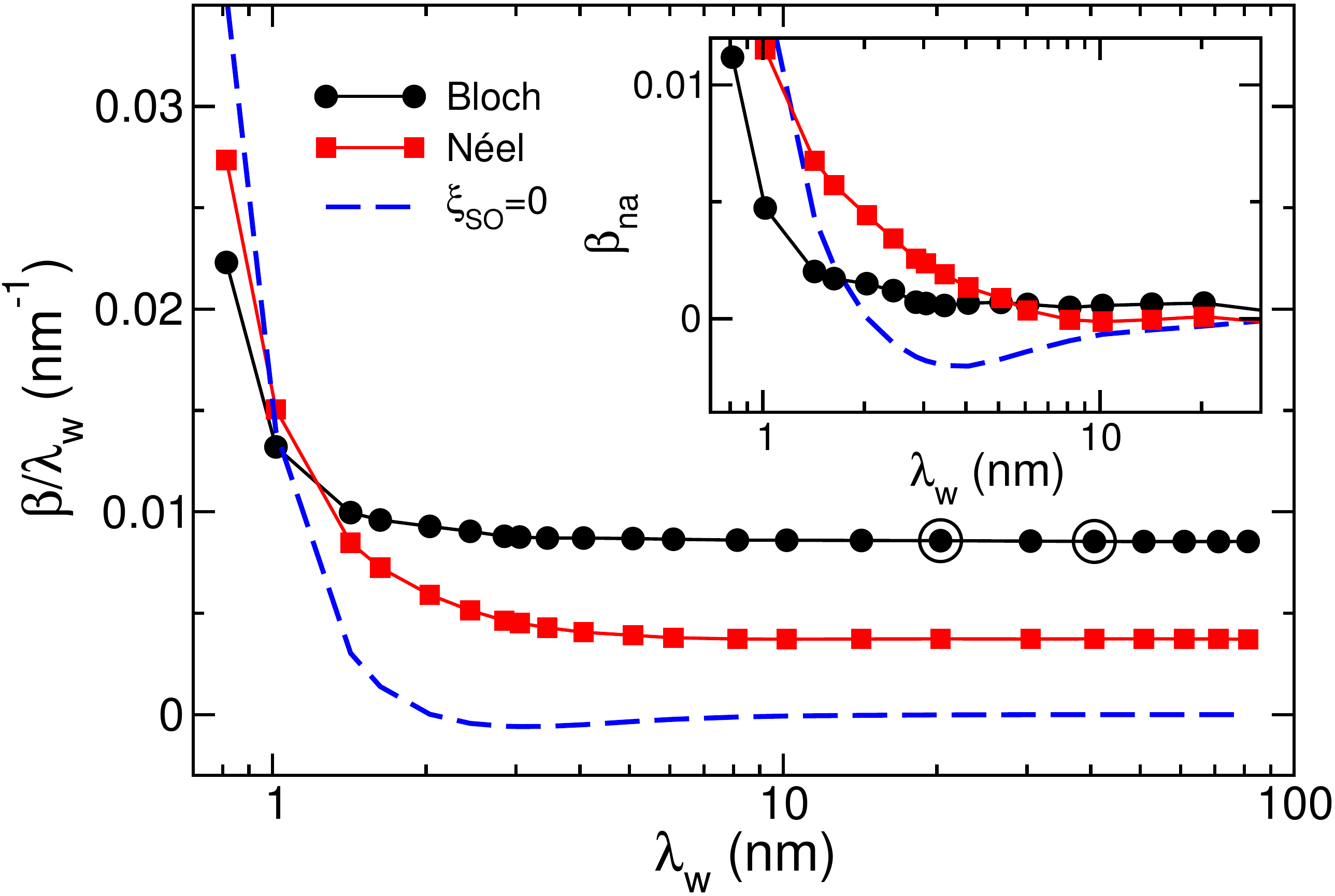}
\caption{Calculated out-of-plane STT parameter $\beta/\lambda_w$ of clean Ni DWs as a function of the DW length $\lambda_w$. $\beta/\lambda_w$ saturates to a finite value for both Bloch and N{\'e}el DWs in the presence of SOC. For short DWs ($\lambda_w<10$~nm), the nonadiabatic contribution to $\beta$ increases dramatically with decreasing DW length; this increase does not depend on SOC. 
The large open circles denotes saturated values of $\beta/\lambda_w=0.0085$ for Bloch DWs with $\lambda_w=$~20 and 40 nm referred to in Sec.~\ref{sec:nickel}.
Inset: the nonadiabatic contribution $\beta_{\rm na}$ as a function of $\lambda_w$, where the contribution proportional to $\lambda_w$ arising from SOC has been subtracted for Bloch and N{\'e}el DWs.
}\label{fig:4}
\end{figure}

The values of $\beta/\lambda_w$ calculated using Eq.~(\ref{eq:beta}) are plotted in Fig.~\ref{fig:4} as a function of the DW length $\lambda_w$. For both Bloch and N{\'e}el DWs, $\beta/\lambda_w$ saturates to a finite value for large values of $\lambda_w$ indicating that $\beta$ diverges in this adiabatic limit. The contribution that is proportional to the DW length arises from SOC; it vanishes if the SOC is switched off in the calculations as shown by the dashed blue line. 

For short DWs with $\lambda_w<10$~nm, there is another contribution to $\beta$ coming from the nonadiabatic reflection of conduction electrons that is not intrinsically related to SOC. This nonadiabatic contribution, $\beta_{\rm na}$, is plotted in the inset to Fig.~\ref{fig:4}, together with the values for Bloch and N{\'e}el DWs with the (SOC-induced) contributions proportional to $\lambda_w$ subtracted. $\beta_{\rm na}$ increases rapidly with decreasing DW length and exhibits oscillations at small values of $\lambda_w$. This nonadiabatic contribution to $\beta$ has been theoretically predicted  and interpreted in terms of standing waves that result from the interference of incoming and reflected electrons \cite{Xiao:prb06, Bohlens:prl10}. 

The divergent contribution arising from SOC is counterintuitive and has not been discussed in the literature. The remainder of this paper will be devoted to  understanding it. To do so, we will use calculations based upon the physically transparent NEM scheme.

\section{Nonequilibrium magnetization scheme\label{sec:nems}}
We begin this section with a brief description of the NEM scheme proposed by N{\'u}{\~ n}ez and MacDonald \cite{Nunez:ssc06, Haney:jmmm08} that can be used to calculate the spatially resolved STT $\bm{\tau}(\mathbf r)$, and of our MTO implementation of this scheme. We illustrate it with calculations for a spin valve consisting of Co and Cu multilayers where, in the absence of SOC, the calculated STT is in good quantitative agreement with the values obtained using a method based upon spin conservation \cite{Wang:prb08}.

\subsection{Formalism}
In the NEM scheme, the torque exerted on a local magnetization ${\bf M}({\bf r})$ is given by
\begin{eqnarray}
\bm{\tau}(\mathbf r)=-\gamma {\bf M}({\bf r})\times {\bf h}^{\mathrm{ex}}({\bf r}),\label{eq:torque1}
\end{eqnarray}
where ${\bf h}^{\rm ex}({\bf r})$ is the exchange field generated by the nonequilibrium magnetization ${\bf m}^{\rm ne}(\bf r)$ induced by a charge current. All occupied states contribute to $\bf M(\mathbf r)$ so that direct calculation of Eq.~(\ref{eq:torque1}) involves an integration over energy up to the Fermi energy. Since an equal and opposite torque is exerted on ${\bf m}^{\rm ne}({\bf r})$ by the local magnetization ${\bf M}({\bf r})$, it can be expressed as 
\begin{eqnarray}
\bm{\tau}(\mathbf r) = -\gamma {\bf H}^{\rm ex}({\bf r}) \times {\bf m}^{\rm ne}({\bf r}), \label{eq:torque2}
\end{eqnarray}
where ${\bf H}^{\rm ex}({\bf r})$ is the exchange field generated by the local magnetization ${\bf M}({\bf r})$\cite{Nunez:ssc06, Haney:prb07, Heiliger:jap08}. Within linear response, ${\bf m}^{\rm ne}({\bf r})$ is composed of contributions from propagating electronic states at the Fermi level. ${\bf H}^{\rm ex}({\bf r})$ only depends on the equilibrium magnetization $\bf M(\mathbf r)$ and can be readily evaluated when carrying out the self consistent equilibrium calculations that involve calculating all occupied states. 
Within the ASA, evaluation of the torque can be simplified by expanding ${\bf H}^{\rm ex}({\bf r})$ and ${\bf m}^{\rm ne}({\bf r})$ in spherical harmonics $Y_{lm}(\bf \hat{r})$ on site ${\bf R}$. On integrating over ${\bf r}$, we find that the torque can be decomposed into site (${\bf R}$) and angular momentum ($l$) resolved contributions as 
\begin{eqnarray}
\bm{\tau}_{\bf R}=\sum_l\bm{\tau}_{{\bf R}l} =-\gamma\sum_l {\bf H}^{\rm ex}_{{\bf R}l} \times {\bf m}^{\rm ne}_{{\bf R}l}. \label{eq:torque3}
\end{eqnarray}

Assuming that the bias $V_b$ applied over the scattering region is infinitesimal, ${\bf m}^{\rm ne}_{{\bf R}l}$ can be constructed from wave functions with energy equal to the Fermi energy 
\begin{eqnarray}
{\bf m}^{\rm ne}_{{\bf R}l} = -\frac{\mu_B}{N_{k_\|}} \sum_{\bf k_\|} 
  \left(
  \sum_{i\in{\mathcal L}} \sum_m \langle \Psi^{i{\bf k_\|}}_{{\bf R}lm}
  \vert \hat{\bm\sigma} \vert \Psi^{i{\bf k_\|}}_{{\bf R}lm} \rangle \right. \nonumber\\
  \left.-
  \sum_{j\in\mathcal R} \sum_m \langle \Psi^{j{\bf k_\|}}_{{\bf R}lm}
  \vert \hat{\bm\sigma} \vert \Psi^{j{\bf k_\|}}_{{\bf R}lm} \rangle  
  \right)   \frac{eV_b}{2}, 
\label{eq:mne}
\end{eqnarray}
where $\Psi^{i{\bf k_\|}}_{{\bf R}lm}$ and $\Psi^{j{\bf k_\|}}_{{\bf R}lm}$ are $lm$ components of the flux-normalized scattering wave functions  ($\vert\Psi\vert^2$ having the dimensions of an inverse energy) with transverse crystal momentum ${\bf k_\|}$, on site ${\bf R}$, incident from the left ($i\in\mathcal L$) and right ($j\in\mathcal R$) leads, respectively. Equation~(\ref{eq:mne}) implies that we consider both right-going electrons from the left lead and left-going holes from the right lead simultaneously.
\footnote{Applying the bias to the chemical potential on the left (right) electrode would lead to an artificial accumulation of electrons (holes) in the scattering region, which should not exist in metallic systems with strong screening. Neglecting such screening does not change the conductance but yields spurious results for local quantities like the chemical potential, nonequilibrium magnetization, spin torque etc. In ballistic systems, the artificial accumulation of charge is usually minimized by applying the bias to the chemical potentials of the left and right lead as $\varepsilon_{\rm F}+eV_b/2$ and $\varepsilon_{\rm F}-eV_b/2$, respectively, as we do in this paper. A more rigorous treatment of screening is necessary for diffusive systems, in particular when SOC is included (R.J.H. Wesselink, Z. Yuan, Y. Liu, A. Brataas, and P.J. Kelly, unpublished); it has negligible effect on the results presented in this paper.} 
Note that the bias $V_b$ in Eq.~(\ref{eq:mne}) will be eventually removed by calculating the torque per unit current density $\bm{\tau}/j$, in units of $\mu_B/(e~\mathrm{nm})$, where $j=GV_b/A$ with $A$ being the cross sectional area. 

The exchange field on site ${\bf R}$ can be decomposed in a similar fashion and ${\bf H}^{\rm ex}_{{\bf R}l}(r)$ obtained by considering test electrons at the Fermi level with up and down spin \cite{Gunnarsson:jpf76}, 
\begin{eqnarray}
{\bf H}^{\mathrm{ex}}_{{\bf R}l} &=& -\frac{\hat{\mathbf M}}{4\mu_B}  
\int dr\left\{ r^2 \left[\upsilon_{{\bf R}\uparrow}(r) - \upsilon_{{\bf R}\downarrow}(r)\right]\right.\nonumber\\
&&\left.\times
\left[
\phi^2_{{\bf R}l\uparrow}  (\varepsilon_{\rm F},r) +
\phi^2_{{\bf R}l\downarrow}(\varepsilon_{\rm F},r)
             \right]\right\}. 
\label{eq:exf}
\end{eqnarray}
Here the radial integration is carried out inside the atomic sphere on site ${\bf R}$ and the $l$-th partial wave $\phi_{{\bf R}l\sigma}(\varepsilon_{\rm F},r)$ is obtained by solving the scalar-relativistic radial equation \cite{Koelling:jpc77} at the energy $\varepsilon = \varepsilon_{\rm F}$ for the spin-dependent radial potential $\upsilon_{{\bf R}\sigma}(r)$ with $\sigma = \uparrow, \downarrow$ \cite{Andersen:prb75}.

\subsection{Test case: Co$|$Cu spin valve}

\begin{figure}[t]
\includegraphics[width=0.95\columnwidth]{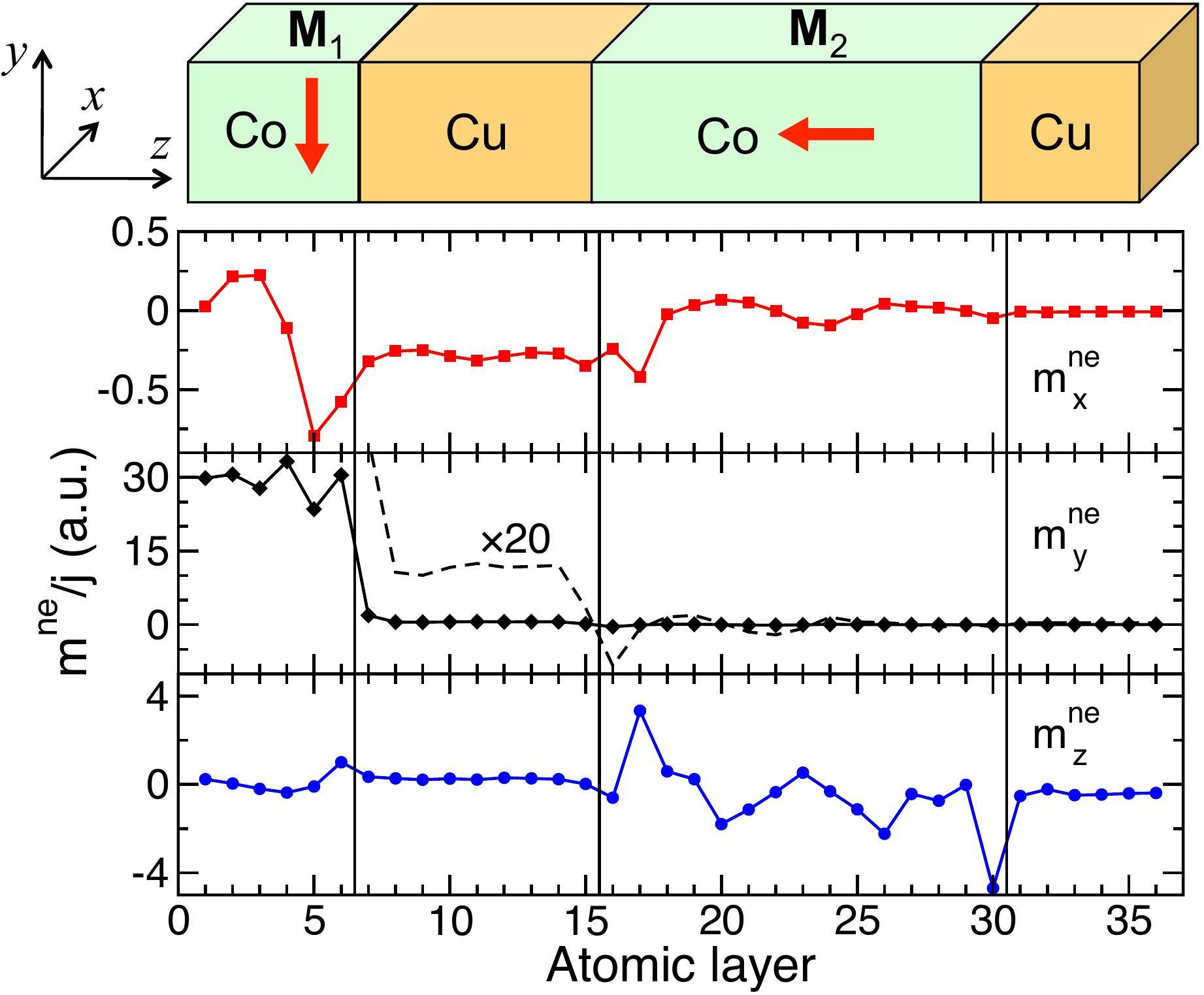}
\caption{Nonequilibrium magnetization calculated without SOC for the [111] oriented  spin valve consisting of 9 layers of fcc Cu and 15 layers of Co sandwiched between semiinfinite Co and Cu leads shown schematically at the top. The dashed line in the central panel is increased by a factor of 20 for clarity.}\label{fig:5}
\end{figure}

To verify our implementation of the NEM scheme, we consider the STTs in a system that has been studied before (without SOC) \cite{Zwierzycki:prb05, Haney:prb07, Wang:prb08}, a Co$|$Cu$|$Co spin valve for which the spin torque has been calculated by assuming spin conservation. The spin valve is schematically shown at the top of Fig.~\ref{fig:5}. The scattering region consists of Co(6)$|$Cu(9)$|$Co(15)$|$Cu(6) with the thicknesses in brackets given in numbers of atomic layers. The left and right leads are bulk Co and Cu, respectively. (In the piecewise self-consistent equilibrium calculations, the atomic sphere potentials in 6 layers of Co on the left and of Cu on the right are allowed to differ from the bulk potentials of the semiinfinite leads.) A uniform lattice constant of 3.55~{\AA} is used and transport is along the fcc [111] with electron flow from left to right. The magnetization directions of the two ferromagnetic Co layers are chosen to be perpendicular to one other, as indicated by the thick arrows in Fig.~\ref{fig:5}. A 2400$\times$2400 $k$ sampling of the 2D Brillouin zone is used to obtain a well-converged out-of-plane component of the torque \cite{Heiliger:jap08}; see the inset to Fig~\ref{fig:7}. SOC was turned off in this test case to compare the results with those obtained with the spin conservation method. \footnote{The results shown in Fig.~5 of Ref.~\onlinecite{Haney:prb07} are for transport along a (001) oriented system. Our results can be compared in detail to those shown in Fig.~4(a) of Ref.~\onlinecite{Wang:prb08}}

\begin{figure}[b]
\includegraphics[width=0.95\columnwidth]{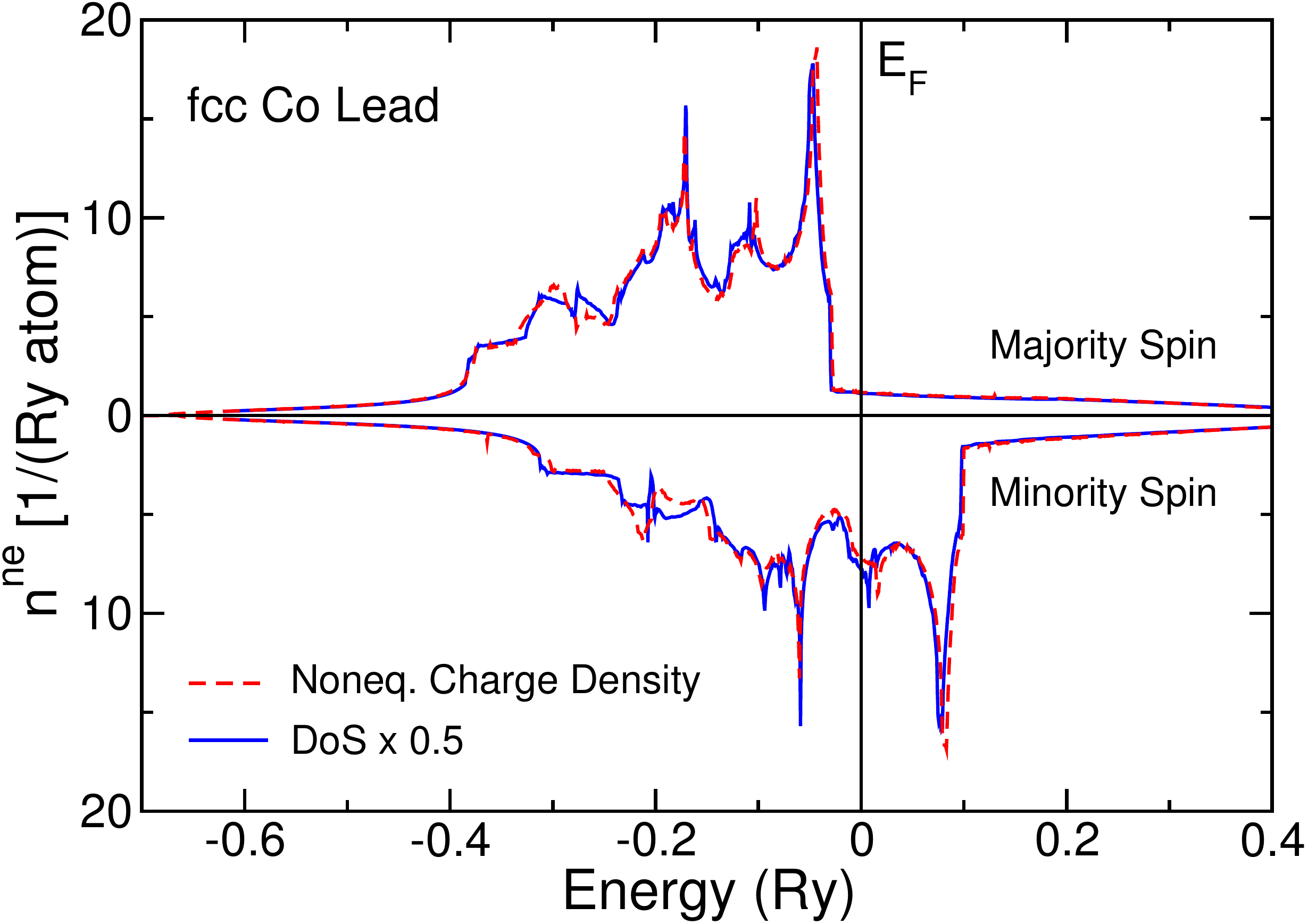}
 \caption{Nonequilibrium spin density of a perfectly crystalline fcc Co lead calculated as a function of the energy. The density of states of bulk fcc Co is plotted for comparison. }\label{fig:6}
\end{figure}

The nonequilibrium magnetization arises from the difference between the nonequilibrium spin densities 
\begin{equation}
n^{\rm ne}_{\mathbf R\sigma}(\varepsilon)=\frac{1}{N_{k_\|}} \sum_{\bf k_\|} 
  \sum_{i\in{\mathcal L}}
  \sum_{l,m} \langle \Psi^{i{\bf k_\|}}_{{\bf R}lm\sigma}(\varepsilon)
  \vert \Psi^{i{\bf k_\|}}_{{\bf R}lm\sigma}(\varepsilon) \rangle.
\label{eq:nne}
\end{equation}
For a perfectly crystalline fcc Co lead, we plot $n^{\rm ne}_{\mathbf R\sigma}$ as a function of the energy $\varepsilon$ of the incoming electrons in Fig.~\ref{fig:6}. The nonequilibrium spin densities equal half of the total density of states (DoS) at the same energy because only electrons incident from the left are considered in Eq.~\eqref{eq:nne} (holes incident from the right contribute the same amount; see Eq.~\eqref{eq:mne}). The muffin tin orbitals used to calculate the DoS with the ``bulk'' LMTO code \cite{Andersen:prb86} are linearized in energy with $\varepsilon_{\nu l\sigma}$ fixed at the corresponding centers of gravity $\overline{\varepsilon}_{\nu l\sigma}$ of the $s$, $p$ and $d$ channels while the DoS is calculated using the tetrahedron method \cite{Blochl:prb94a}. The nonequilibrium spin density $n^{\rm ne}_{\bf R\sigma}(\varepsilon)$ on the other hand is obtained in the scattering code with $\varepsilon_{\nu l\sigma}=\varepsilon$ and with discrete summation over ${\bf k}_{\|}$. These factors account for the slight differences seen in Fig.~\ref{fig:6}. At the Fermi level, minority spins contribute more nonequilibrium states so ${\bf m}^{\rm ne}$ is antiparallel to the local magnetization $\bf M$ in Co that is dominated by the occupied majority spin states.

The nonequilibrium magnetization $\mathbf m^{\rm ne}$ generated in the Co(6)$|$Cu(9)$|$Co(15)$|$Cu(6) spin valve by the electric current is plotted in Fig.~\ref{fig:5}. Since there is no disorder in either the Cu or Co layers, scattering only occurs at the interfaces. At a Co$|$Cu interface, there is a large mismatch between the Cu and Co electronic structures for the minority spin channel leading to a significant reflection of these electrons. This corresponds to a large minority-spin interface resistance \cite{Xia:prb01, Xia:prb06} and leads to the accumulation of the minority spin density $m_y^{\rm ne}$ seen in Fig.~\ref{fig:5} antiparallel to the local magnetization direction in layer M$_1$. The oscillations (between layers 0 and 7) are a consequence of the interference between incident and reflected waves. 

The magnetizations of the two ferromagnetic layers in Fig.~\ref{fig:5} are perpendicular to each other. The spin current transmitted through the first ${\bf M}_1|$Cu interface is oriented along the $-y$ direction. In this Cu ``spacer'' layer, accumulation of nonequilibrium magnetization is mostly of minority-spin electrons (along $+y$) injected through the Co$|$Cu interface. There are also contributions (along $-x$) from multiple scattering at the two Cu interfaces. Without a local magnetization (and spin relaxation) in Cu, these propagating states keep their spin polarization. The quantization axis of the ${\bf M}_2$ layer is at right angles, along the $-z$ direction. Spins injected from the left, oriented perpendicular to this quantization axis, precess in ${\bf M}_2$. This results in the oscillatory behavior seen for ${\rm m}^{\rm ne}_x$ and ${\rm m}^{\rm ne}_y$ in the ${\bf M}_2$ layer in Fig.~\ref{fig:5}. In addition, components of $\mathbf m^{\rm ne}$ transverse to ${\bf M}_2$ decay into the ferromagnetic layer as a result of dephasing \cite{Zwierzycki:prb05}. In ferromagnetic Co, the transverse components of $\mathbf m^{\rm ne}$ vanish after propagating about 3 nm (15 atomic layers) \cite{Zwierzycki:prb05, Haney:prb07, Wang:prb08}. Eventually, the longitudinal components have the largest magnitude in the ferromagnetic layers, i.e. $m_y^{\rm ne}$ in the left Co layer and $m_z^{\rm ne}$ in the right one.

\begin{figure}[t]
\includegraphics[width=0.95\columnwidth]{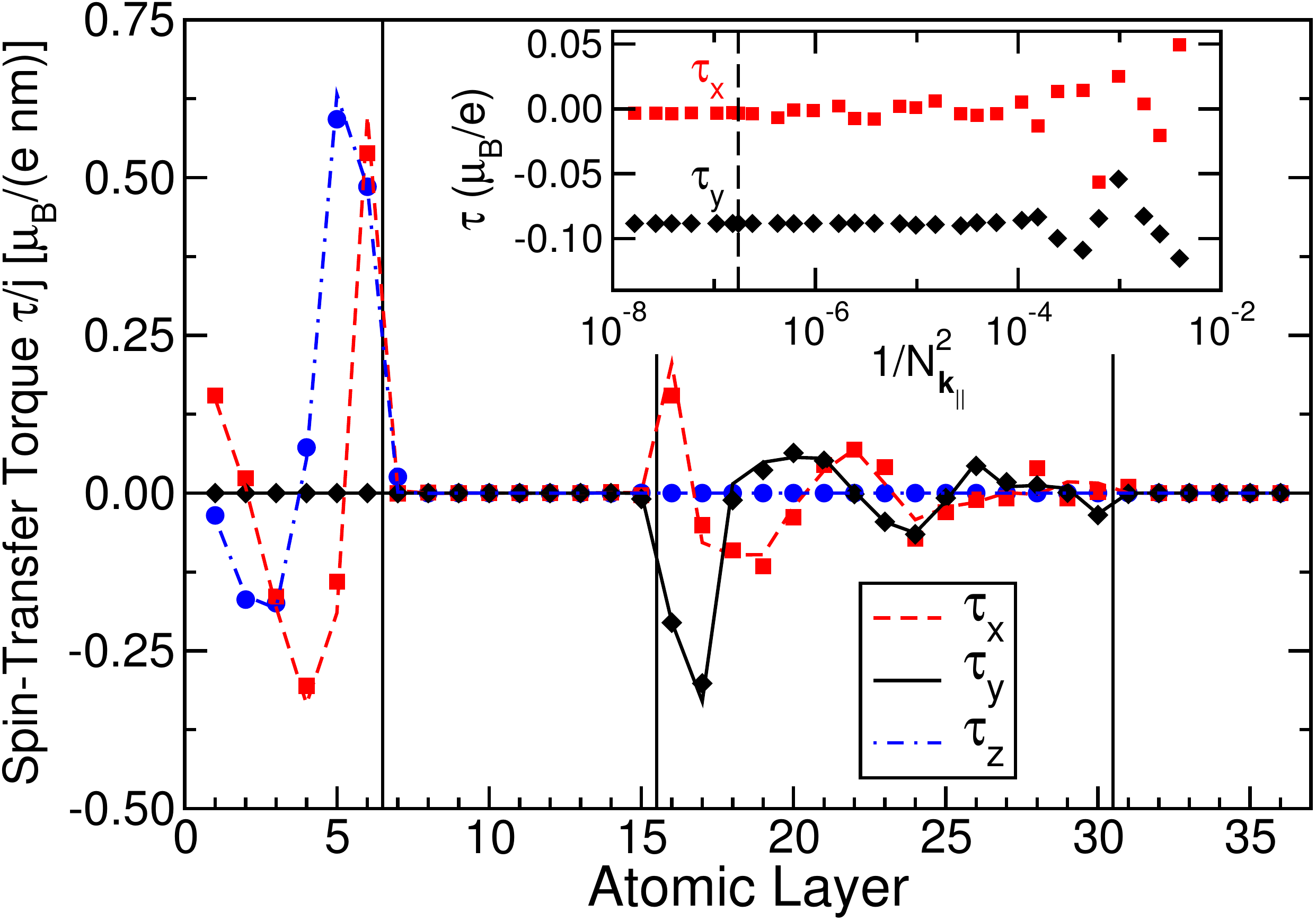}
\caption{STT calculated for a spin valve consisting of Co and Cu multilayers without SOC. The lines are calculated based on spin conservation method\cite{Wang:prb08} while the symbols are obtained using the NEM scheme. Inset: total STT on $\mathbf M_2$ as a function of the $k$ sampling density. The vertical dashed line indicates the final sampling density (2400$\times$2400) adopted. 
}\label{fig:7}
\end{figure}

In spite of their large magnitudes, the longitudinal components of $\bf m^{\rm ne}$ do not exert torques on the local magnetization; the smaller transverse components do. The spin torques ${\bm \tau}$ calculated using Eqs.~\eqref{eq:torque3}, \eqref{eq:mne}, and \eqref{eq:exf} are plotted as a function of position in Fig.~\ref{fig:7}. Reflecting the oscillations in $\mathbf m^{\rm ne}$, the calculated STTs also display oscillations in the Co ferromagnetic layers. The total STT 
\begin{equation}
{\bm \tau}_{{\bf M}_2} =\sum_{{\bf R}\in {{\bf M}_2}} {\bm \tau}_{\bf R}
\end{equation}
exerted on the right Co layer ($\mathbf M_2$) is plotted in the inset to Fig.~\ref{fig:7}. It has a large in-plane component in the $-y$ direction and one order of magnitude smaller out-of-plane component in the $-x$ direction. The feature agrees with the spin-transfer picture \cite{Slonczewski:jmmm96, Berger:prb96} where the conduction electrons polarized by $\mathbf M_1$ transfer their spin angular momentum to $\mathbf M_2$ resulting in a STT parallel to $\mathbf M_1$. Finally, the STTs calculated using the NEM scheme and spin conservation method are in perfect mutual agreement (within the numerical accuracy) and in good agreement \cite{Note2} with earlier NEM \cite{Haney:prb07} and spin conservation \cite{Wang:prb08} calculations. 

\section{SOC-induced STTs in ballistic Ni DWs\label{sec:nickel}}
\begin{figure}[t]
\includegraphics[width=0.95\columnwidth]{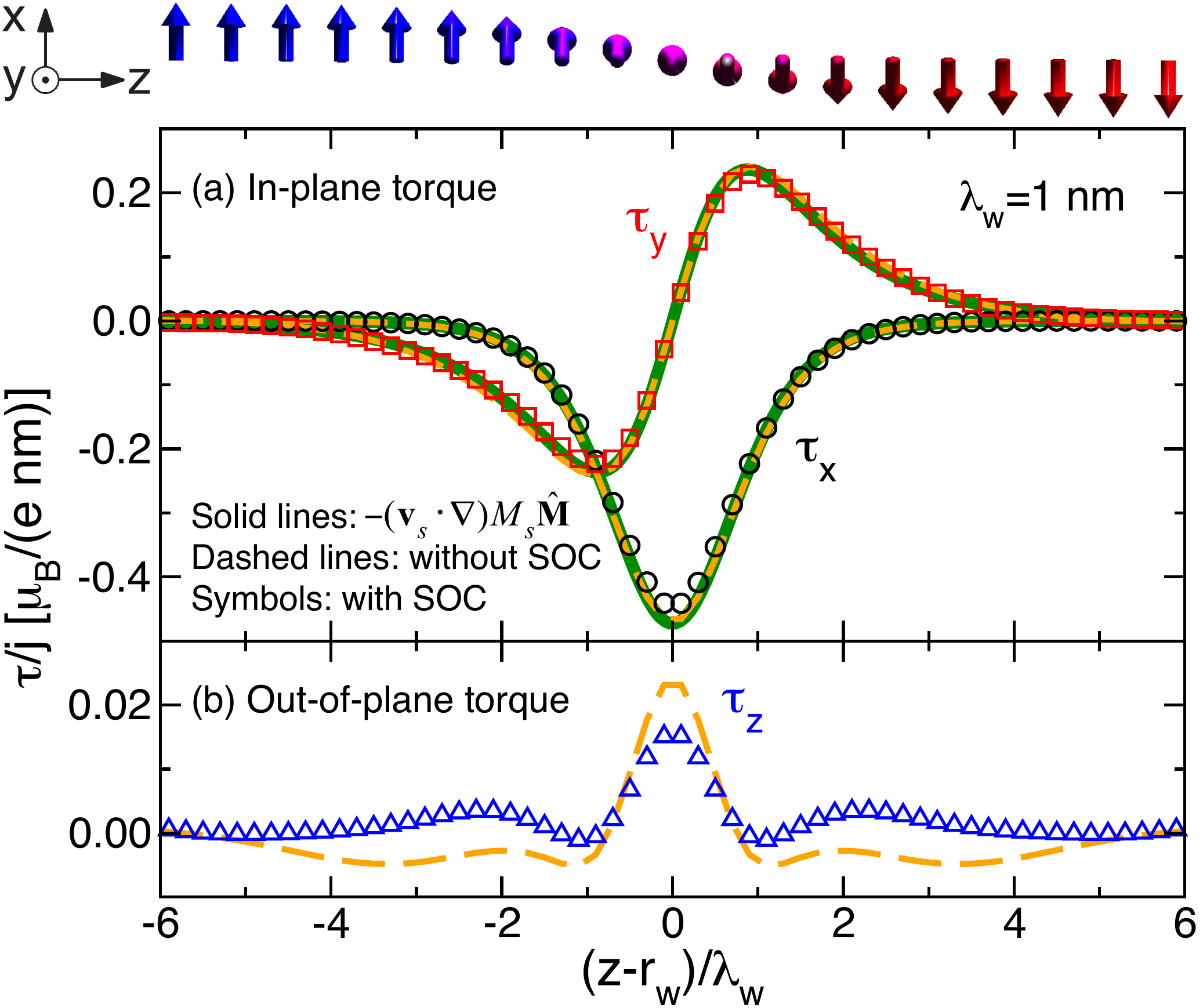}
\caption{Calculated STT for a short, clean Ni Bloch DW with $\lambda_w=1$~nm. The thick solid (green) lines in (a) show the adiabatic form of in-plane torque $-(\mathbf v_s\cdot\nabla)M_s\hat{\mathbf M}$. The dashed (orange) lines in (a) and (b) are the STTs calculated without SOC and the symbols are obtained with SOC. Including SOC gives only slight changes in the calculated STTs. The out-of-plane torque $\tau_z$ mainly results from the abrupt variation of the exchange potential in the center of the short DW. }\label{fig:8}
\end{figure}

In this section, we apply the NEM scheme to calculate the spatially resolved STT for Bloch DWs (see Fig.~\ref{fig:8} for the profile) in ballistic Ni in order to obtain a transparent physical picture of the interplay between an electrical current and local magnetization that results from SOC. In particular, we wish to understand the unexpected divergence of $\beta$ found in the adiabatic limit with the charge pumping formalism. The numerical details are the same as described in Sec.~\ref{sec:numerical} except that a denser $k$ mesh of 2400$\times$2400 points is used to sample the two-dimensional Brillouin zone.

In the generalized LLG equation, Eq.~(\ref{eq:llg}), the expression $-(\mathbf v_s\cdot\nabla)\hat{\mathbf M}$ for the in-plane torque comes from spin conservation. In deriving it, it was assumed that conduction electrons can adiabatically follows the orientation of the local magnetization \cite{}. At position $\mathbf r$, the spin current carried by an electrical current $j$ is given by $\hbar\gamma Pj \hat{\mathbf M}(\mathbf r)/(2e)$ and the loss of spin current a short distance away from $\mathbf r$ corresponds to the STT $-(\mathbf v_s\cdot\nabla)M_s\hat{\mathbf M}(\mathbf r)$. Using the analytical expression for the one-dimensional magnetization profile, the absolute magnitude of the adiabatic torque $-\mathbf v_s M_s d\hat{\mathbf M}(z)/dz$ is
\begin{eqnarray}
\frac{\tau_x(z)}{j}&=&\frac{\mu_B P}{e\lambda_w}\mathrm{sech}^2\frac{z-r_w}{\lambda_w},\label{eq:adia1}\\
\frac{\tau_y(z)}{j}&=&-\frac{\mu_B P}{e\lambda_w}\tanh\frac{z-r_w}{\lambda_w}\mathrm{sech}\frac{z-r_w}{\lambda_w}.\label{eq:adia2}
\end{eqnarray}
This adiabatic torque is plotted in Fig.~\ref{fig:8}(a) as solid green lines for a very short DW with $\lambda_w=1$~nm, where $P=(G^{\uparrow}_{\rm Sh}-G^{\downarrow}_{\rm Sh})/(G^{\uparrow}_{\rm Sh}+G^{\downarrow}_{\rm Sh})=-0.48$ is obtained from the spin-resolved Sharvin conductances $G^{\uparrow}_{\rm Sh}$ and $G^{\downarrow}_{\rm Sh}$ without SOC. The negative value of $P$ indicates that the minority-spin channel has more propagating states than the majority-spin channel at the Fermi level; the $s$ band contribution is very similar for both spins while the $d$ contribution is absent from the majority spin channel.

The calculated in-plane STTs shown in Fig.~\ref{fig:8}(a) are seen to accurately follow the adiabatic form regardless of SOC. The near perfect coincidence of the dashed lines (without SOC) and symbols (with SOC) superposed in Fig.~\ref{fig:8}(a) on the thick solid lines indicate that the adiabatic form captures most of the in-plane STT even in such a short DW. This result is in agreement with a previous calculation for free-electron Stoner-model DWs where the deviation of the in-plane STT from the adiabatic torque was found to be very small. \cite{Xiao:prb06} 

The out-of-plane STT, plotted in Fig.~\ref{fig:8}(b), is seen to be mainly localized at the DW center. Since the adiabatic forms in Eqs.~(\ref{eq:adia1}) and (\ref{eq:adia2}) do not have an out-of-plane $(\tau_z)$ component the appearance of such a STT implies a nonadiabaticity of the conduction electrons moving through the DW. In short DWs the out-of-plane torque arises from the nonadiabatic reflection of conduction electrons, especially in the central region of the DW where the magnetization has the largest spatial gradient; including SOC has relatively little effect. These features are consistent with the observation from Fig.~\ref{fig:4} that the calculated out-of-plane parameter $\beta \ll 1$ and is not very sensitive to SOC in short DWs. 

\begin{figure}[b]
\includegraphics[width=0.95\columnwidth]{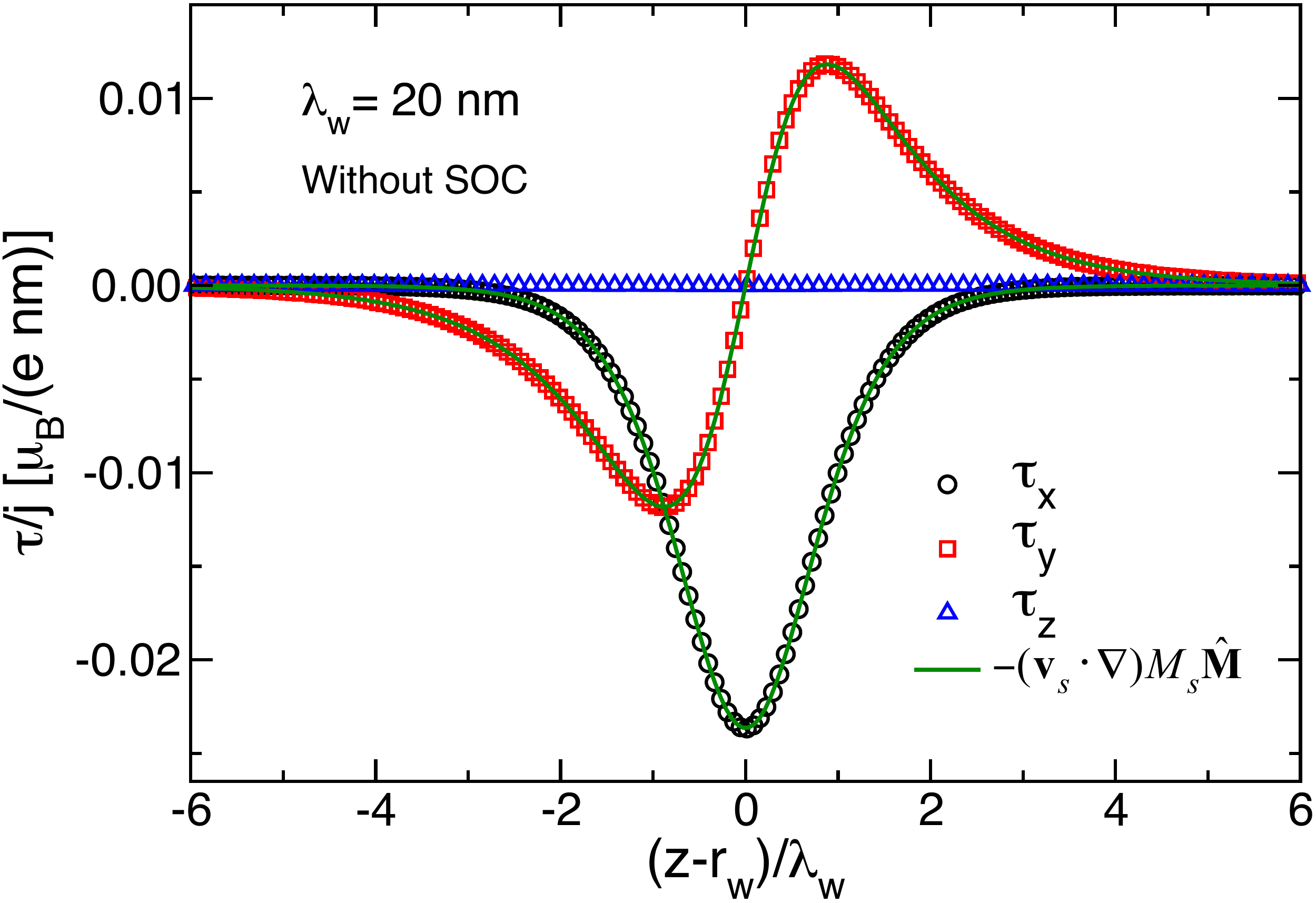}
\caption{STTs calculated without SOC for a long, clean Ni Bloch DW with $\lambda_w=20$~nm. The in-plane torque ($\tau_x$ and $\tau_y$) follows the adiabatic form $-(\mathbf v_s\cdot\nabla)\mathbf M$ resulting from spin conservation. The out-of-plane torque ($\tau_z$) vanishes without SOC in such a long DW.}\label{fig:9}
\end{figure}

Without SOC, the nonadiabatic contribution to the $\beta$ torque observed in short DWs decreases as the magnetization gradients become smaller in longer walls; see Fig.~\ref{fig:9} for $\lambda_w=20$~nm. The in-plane components in this case completely follow the adiabatic form $-(\mathbf v_s\cdot\nabla)M_s\hat{\mathbf M}$ and the out-of-plane STT vanishes within the numerical accuracy. Analysis of the conductance shows that only 0.17\% of incoming electrons from the leads are reflected by this $\lambda_w=20$~nm DW; the others pass through the DW by adjusting their spins adiabatically. 

\begin{figure}[t]
\includegraphics[width=0.95\columnwidth]{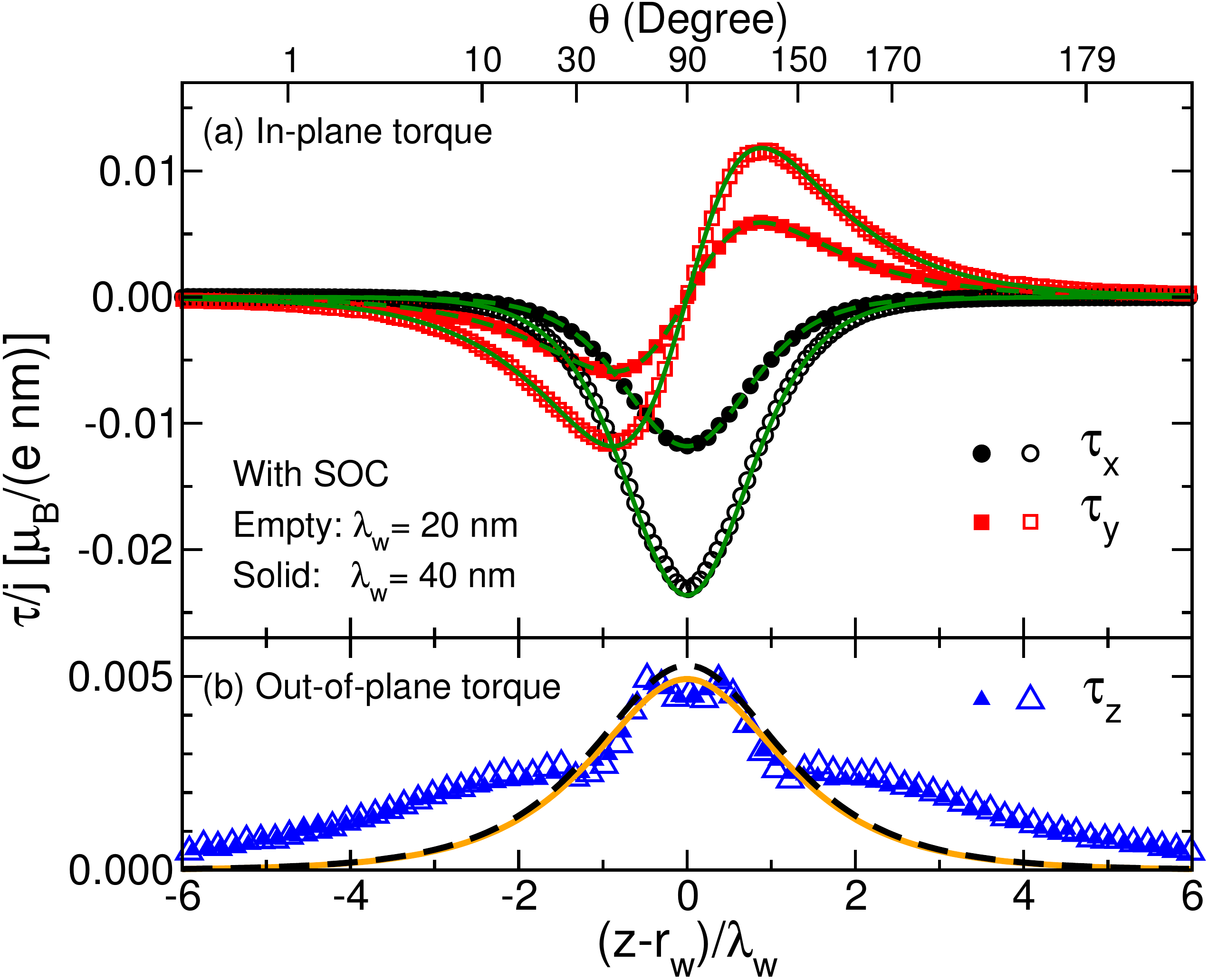}
\caption{STTs calculated with SOC for a long, clean Ni DW with $\lambda_w=20$ (empty symbols) and 40~nm (solid symbols). The in-plane torque (a) is found to be nearly proportional to the magnetization gradient in agreement with the expressions for STT arising from loss of spin current: the solid and dashed lines denote the adiabatic form $-(\mathbf v_s\cdot\nabla)\hat{\mathbf M}$ for $\lambda_w=20$ and 40~nm, respectively. The out-of-plane torque (b) arises from the SOC-induced electron reflection due to conduction channel mismatch and is independent of the magnetization gradient. The dashed black and solid orange lines illustrate unweighted and weighted fits to Eq.~(\ref{eq:fit}), respectively.
}\label{fig:20}
\end{figure}

With SOC included, the electron reflection in long DWs is mainly due to the intrinsic DWR \cite{Nguyen:prl06, Oszwaldowski:prb06} and results in out-of-plane torques. The STTs calculated with SOC in two long DWs ($\lambda_w=20$ and 40~nm) are plotted in Fig.~\ref{fig:20} where the in-plane (a) and out-of-plane (b) components show different length dependences. The in-plane STT is smaller in the longer DW because it results mainly from the adiabatic spin transfer mechanism \cite{Slonczewski:jmmm96,Berger:prb96} and is proportional to the magnetization gradient. Note that we plot the STTs with the scaled coordinates $(z-r_w)/\lambda_w$ so the integral of the in-plane torque with respect to $z$ is always $-2v_s M_s\hat{\mathbf M}(-\infty)$, independent of $\lambda_w$ though the maximum in-plane torque is proportional to $1/\lambda_w$. The reflection of electrons due to conduction channel mismatch contributes very little to the in-plane torques because only a small number (1.8\%) of incoming electrons are reflected resulting in a contribution to $\mathbf m^{\rm ne}$ that is much smaller than that due to the adiabatic spin-transfer mechanism. Therefore the in-plane STTs still follow the adiabatic form as we already saw in Fig.~\ref{fig:9}(a).

The most striking effect of SOC is seen in the out-of-plane torques, which have the same amplitude for both DW lengths at the same scaled position $(z-r_w)/\lambda_w$; see Fig.~\ref{fig:20}(b). Alternatively, and equivalently, we can characterize the scaled position with a winding angle $\theta$ which rotates from 0 to $\pi$ for a 180$^\circ$ DW. The electron reflection arising from conduction channel mismatch is constant and independent of $\lambda_w$ in an arbitrary interval where the magnetization rotates from $\theta$ to $\theta+\Delta\theta$. Consequently the nonequilibrium magnetization $\mathbf m^{\rm ne}$ in this interval is also constant giving rise to STTs with the same amplitude. Because the expression for the out-of-plane torque in Eq.~(\ref{eq:llg}) contains the magnetization gradient (or $1/\lambda_w$), a factor $\lambda_w$ must be included in the parameter $\beta$ in Eq.~(\ref{eq:llg}) to reproduce the NEM result shown in Fig.~\ref{fig:20}(b) of a ``constant'' local out-of-plane torque. This is the reason why the value of $\beta$ calculated with SOC in the adiabatic limit is proportional to $\lambda_w$ in Fig.~\ref{fig:4}. 

To confirm the quantitative agreement between Fig.~\ref{fig:4} and Fig.~\ref{fig:20}(b), we fit the out-of-plane STTs in Fig.~\ref{fig:20}(b) that were calculated numerically with the NEM scheme to the analytical form in Eq.~(\ref{eq:llg}) 
\begin{equation}
\tau_z(z)=\frac{\beta}{\lambda_w}\frac{v_sM_s}{j}\mathrm{sech}\frac{z-r_w}{\lambda_w}.\label{eq:fit}
\end{equation}
If we use a uniform weight in the fitting, we obtain $\beta/\lambda_w=0.009$~nm$^{-1}$ (black dashed line in Fig.~\ref{fig:20}). From the charge pumping calculations we know that displacing the DW rigidly with $r_w$ results in relatively large precession at the center of the DW; the further from the center, the less the magnetization changes. Then we can also fit $\tau_z(z)$ using the $z$ dependent weight $\mathrm{sech}\frac{z-r_w}{\lambda_w}$ to find the fitted value of $\beta/\lambda_w$=0.0086~nm$^{-1}$ (solid orange line in Fig.~\ref{fig:20}), which is in perfect agreement with the value 0.0085~nm$^{-1}$ in Fig.~{\ref{fig:4}} (large open circles) obtained from the charge pumping calculations. 

\section{Conclusions\label{sec:conclusion}}
Using Landauer-B{\"u}ttiker scattering theory combined with first-principles electronic structure calculations, we have implemented two computational schemes capable of describing spin torques in the presence of spin-orbit interaction, namely, the charge pumping \cite{Nunez:ssc06} and the nonequilibrium magnetization \cite{Hals:prl09} formalisms. The charge pumping formalism efficiently determines the total current-induced torque in terms of the charge current pumped by a precessing magnetization. We have used this scheme to calculate the DWR and out-of-plane STT parameter $\beta$ for ballistic nickel DWs. In addition to the nonadiabatic reflection of conduction electrons by the  rapidly varying exchange potentials that leads to a large DWR for very short DWs, an intrinsic DWR arising from SOC dominates the DWR at large DW lengths \cite{Nguyen:prl06, Oszwaldowski:prb06}. When SOC is included, the out-of-plane STT parameter $\beta$ is found to be proportional to the DW length in the adiabatic limit. To understand this unexpected behavior, we implemented the NEM scheme that can be used to calculate position resolved STTs and is physically transparent. We illustrate the NEM scheme using a Co$|$Cu$|$Co spin valve as an example. In particular, without SOC the NEM scheme reproduces the STTs obtained for the spin valve from the spatial variation of the spin current combined with spin conservation. 

Applying the NEM scheme to calculate position-resolved STTs in ballistic Ni DWs, we demonstrate that the in-plane STT can be described by the adiabatic form from the generalized LLG Eq.~(\ref{eq:llg}) for both short and long DWs independent of SOC. The position dependent torques calculated using the NEM scheme allow us to understand the behavior of $\beta$ obtained with the charge pumping formalism. In short DWs the nonadiabatic reflection of conduction electrons is the main reason for the out-of-plane torque, independent of SOC. In the adiabatic limit, the anisotropic distribution of conducting channels resulting from SOC that gives rise to the intrinsic DWR contributes to an out-of-plane torque. This contribution is constant at a given winding angle of a DW such that the parameter $\beta$ in the generalized LLG Eq.~(\ref{eq:llg}) is proportional to the DW length in quantitative agreement with the result of the charge pumping formalism.

\begin{acknowledgments}
We would like to thank Arne Brataas, Kjetil Hals, Yi Liu, Jiang Xiao, Frank Freimuth, Jairo Sinova, Lei Wang and Pengxiang Xu for helpful discussions. This work was financially supported by the ``Nederlandse Organisatie voor Wetenschappelijk Onderzoek'' (NWO) through the research programme of ``Stichting voor Fundamenteel Onderzoek der Materie'' (FOM) and the  supercomputer facilities of NWO ``Exacte Wetenschappen (Physical Sciences)''. It was also partly supported by the Royal Netherlands Academy of Arts and Sciences (KNAW). Z. Y. acknowledges the financial support of the Alexander von Humboldt Foundation.
\end{acknowledgments}

\end{document}